\documentclass[a4paper,11pt]{article}

\usepackage[latin1]{inputenc}
\usepackage{graphicx}
\usepackage[intlimits]{amsmath}

\newtheorem{Ipotesi}{Hypotesis}
\newtheorem{Proposizione}{Proposition}
\newtheorem{Teorema}{Theorem}
\DeclareMathOperator{\diff}{d}

\def\veps{  {\varepsilon} }
\def\interi{\mathinner{\bf N}}
\def\reali{\mathinner{\bf R}}

\def\={\stackrel{\mathrm{def}}{=}}
\def\Zi{\mathcal{Z}_j}

\title{Thermodynamics and time--averages}
\author{ A. Carati\footnote{ Universit\`a di Milano, Dipartimento di Matematica
          Via Saldini 50,  20133 Milano (Italy)  
          E-mail: {\tt carati@mat.unimi.it} }
        }

\begin{document}

\maketitle

\centerline {ABSTRACT} 
\noindent  
\vskip 1truecm
For a dynamical system far from equilibrium, one has to deal with
empirical probabilities defined through time--averages, and the main
problem is then how to formulate an appropriate statistical
thermodynamics. The common answer is that the standard functional
expression of Boltzmann-Gibbs for the entropy should be used, the
empirical probabilities being substituted for the Gibbs measure.
Other functional expressions have been suggested, but apparently with
no clear mechanical foundation. Here it is shown how a natural extension
of the original procedure employed by Gibbs and Khinchin in defining
entropy, with the only proviso of using the empirical probabilities,
leads for the entropy to a functional expression which is in general different
from that of Boltzmann--Gibbs.  In particular, the Gibbs entropy is
recovered for empirical probabilities of Poisson type, while the Tsallis
entropies are recovered for a deformation of the Poisson distribution.
\vfill  
\noindent
Running title: Thermodynamics and time--averages  
\vskip 3.truecm
\noindent
\eject 

\section{Introduction} 
Classical equilibrium statistical mechanics is a well established
subject. One deals with a dynamical system defined by a Hamiltonian 
function $H$ on a phase space $\mathcal{M}$, which is provided with a suitable
probability measure (typically Lebesgue measure). Having fixed a value $U$
of the mean energy, one then correspondingly defines a conditional probability
(the Gibbs measure) on $\mathcal{M}$. To obtain statistical
thermodynamics, the classical procedure of Gibbs and Khinchin (see \cite{khinchin})
consists first of all in identifying
the external work  $\delta \mathcal{W}$ as
$\delta \mathcal{W}=<\partial_{\kappa} H > \diff \kappa $, where $<\cdot>$
denotes average with respect to Gibbs measure, and the Hamiltonian has
been assumed to depend on an external parameter $\kappa$.
A corresponding expression for the exchanged heat $\delta Q$ is thus
obtained, as $\delta Q=\diff U -\delta \mathcal{W}$, and this 
finally allows to introduce  entropy in the
standard thermodynamic way.

However one often deals with situations in which the relevant probability
measure is different from that of Gibbs. Typically this occurs when
the probability is defined dynamically in terms of time--averages
(sojourn times) and the final time is not long enough for equilibrium
to have been attained (statistical mechanics far from equilibrium).
In such a case, following  Poincar\'e, Boltzmann and
Einstein one can assume  that the expectations should be computed in
terms of  time--averages.

The problem is then to determine which is  
the correct  expression for the thermodynamic functions in such a
situation,  and for
this it is sufficient to provide an expression for the entropy.
Many people accept the thesis that  entropy should be defined by
the classical formula of Gibbs with the only proviso that the
empirical probabilities (time--averages) should be substituted for the
Gibbs measure. Other people, particularly Tsallis, suggest instead
that different formul\ae\ should be used, but a mechanical foundation
for such formul\ae\ is apparently lacking.

In the present paper we show how the standard procedure to define
entropy recalled above can be implemented also if, in the expression
for the external work, the averaging is performed through the
empirical probabilities rather than through Gibbs measure.  In
particular, the standard Gibbs entropy is obtained for empirical
probabilities of Poisson type, while the Tsallis entropies are
recovered for a deformation of the Poisson distribution.  The
implementation of the Gibbs--Khinchin procedure requires however to
previously define a suitable mathematical setting for dealing with the
empirical probabilities.  This is dealt with in the first two
sections. The thermodynamics is then developed, and the two cases
leading to the Gibbs and the Tsallis distributions are finally
considered. Some open problems are also mentioned.  The style of the
paper will be quite informal, with no striving for strict mathematical
rigour.
\section{ A priori probability}
Suppose one has a map $\phi : \mathcal{M}\to\mathcal{M}$ (for example
the time--flow of an autonomous Hamiltonian system at time $\Delta t$) 
in a given phase space $\mathcal{M}$, and
we are interested in computing time--averages of a dynamical variable
$A(x)$ (a real function on $\mathcal{M}$):
$$
\bar A (x_0) {\=} \frac 1N  \sum_{n=1}^N A(x_n) \qquad \hbox{for}\quad
N\gg1 \ ,
$$
the sequence $\{x_n\}$ being defined by the recurrence
$x_{n+1}=\phi(x_n)$. 
The number $N$ thus plays the role of the ``final'' time, and since
now on will be thought of as a fixed parameter.
One can divide the space $\mathcal{M}$ into a large number $K$ of
disjoint cells $\Zi$ (such that $\mathcal{M}=\cup \Zi$), and one has
then 
$$
\bar A(x_0) \simeq \sum_{j=1}^K A_j \frac {n_j}N \ ,
$$
where $A_j$ is the value of $A$ in a point $x\in\Zi$, and $n_j$ is the
number of times the sequence $\{x_n\}$ visits $\Zi$. It is clear that
$n_j$ depends on $x_0$ so that, if a certain probability distribution is
assigned for the initial data $x_0$, correspondingly $n_j$ turns out
to be a random variable with a certain distribution function $F_j(n)$,
which will depend both on the dynamics (i.e. on the map $\phi$) and on
the distribution of the initial data. So one can speak in general of
the ``a priori probability $P$ that the cell $\Zi$ will be visited a
number of times $n_j\le n$:''
\begin{equation}\label{eq1}
P(n_j \le n)=F_j(n) \ .
\end{equation}
In the following, in order to simplify the discussion, it will be  
supposed that
such a probability \emph{does not depend } on the cell $\Zi$
(i.e. $F_j(n)=F(n)$ $\forall j$). The general case can also be easily
dealt with. With an abuse of notation we will denote by $F(n_j)$ the
probability that the cell $\Zi$ is visited a number of times $\le n_j$,
following the common attitude of using the same letter both for a
random variable and for its value.

A central point in the discussion is that the time--average $\bar
A(x_0)$ is itself a random variable, so that it is meaningful to
consider its expectation. Denoting by $<\cdot>$ expectation with
respect to the a priori distribution, one has then
$$
<\bar A> = \frac 1N \sum_{j=1}^K A_j <n_j> \ . 
$$

A step forward is made if one supposes that the expectation of $n_j$
can be computed by a formalism analogous to the Grand Canonical one,
i.e. if one introduces the
\begin{Ipotesi}
The quantities $n_j$ are independent random variables,
conditioned by $\sum n_j=N$.
\end{Ipotesi}
Using this hypothesis, the expectation of $\bar A$ can be computed as
\begin{eqnarray}
<\bar A > & = & \frac 1N \sum_j A_j <n_j> =\frac 1N \sum_j A_j \frac {
     {\idotsint}_{\sum n_i=N} \, n_j 
                  \diff F(n_1)\cdots \diff F(n_K) }
    {\idotsint_{\sum n_i=N} \diff F(n_1)\cdots \diff F(n_K)} \nonumber \\ 
 \qquad   & = & \frac 1N  \frac {
     \idotsint_{\sum n_i=N}{\sum\displaylimits}_j  n_j A_j \diff F(n_1)\cdots \diff F(n_K) }
    {\idotsint_{\sum n_i=N} \diff F(n_1)\cdots \diff F(n_K)}  \ .
\end{eqnarray}
In particular, introducing the function
$$
Z(\lambda,A)\= \idotsint_{\sum n_j=N} e^{-\lambda \frac 1N \sum_{j} n_j A_j } 
         \diff F(n_1)\cdots \diff F(n_K) \ ,
$$
which generates the moments of the random variable $A$, one obtains
\begin{equation}
 <\bar A > = - \left. \frac 1N \partial_{\lambda}
             \log Z(\lambda,A)\right|_{\lambda=0} \ .
\end{equation}

\section{Conditional Probability and Large Deviations}
Usually in statistical thermodynamics  one does not  deal directly
with the a priori probability,
because it is generally assumed that the time--average of 
a certain macroscopic quantity, typically the total energy,
has a given value, which should play
the role of an independent variable. So we consider the total energy,
which we now denote by $\veps$, and its time--average $\bar \veps
=1/N\sum \veps_j n_j$, and we impose on the numbers $n_1, \cdots, n_K$
the condition
$$
 \frac 1N \sum_{j=1}^K \veps_j n_j = U  = \mathrm{const} \ .
$$
Actually in such a way one meets with
a  large deviation  problem, 
because one usually also assumes that one has
$$
U\ne < \bar \veps > =\frac 1N \sum_{j=1}^K \veps_j <n_j> \ ,
$$
and furthermore
that $U\, - <\bar \veps>$ is large, against 
the law of large numbers.
One is thus in an extremely unlikely situation,  where the
information on the value $U$ of $\bar \veps$ 
turns out to have  a great relevance, by conditioning
the expectations  of the other quantities of interest.
The problem we discuss in the rest of the present 
section is indeed  how to compute
such a conditional probability, or
\emph{a posteriori probability},  as it is also called.

Using the tool of the  moment function, the a posteriori expectation
of $\bar A$, denoted by $<\bar A >_{U}$, is given by
\begin{equation}\label{eq2}
<\bar A >_{U}= - \left. \frac 1N \partial_{\lambda}
              \log Z(\lambda,A,U)\right|_{\lambda=0} , 
\end{equation}
where the moment function $Z$ is defined by

\begin{multline}
  Z(\lambda,A,U) \= \int_{\reali} \diff A\, \exp(-\lambda A) 
                  \idotsint\diff F(n_1)\cdots \diff F(n_K) \\
  \delta(A-\frac 1N\sum A_j n_j)  \delta(U-\frac 1N\sum \veps_j n_j)  
  \delta(N - \sum n_j)             \ ,         
\end{multline}
$\delta(x)$ being the usual Dirac's function. Here, use was made of the
familiar formula for the conditional probability, namely the formula
$P(\mathcal(A)|\mathcal(B)) =
P(\mathcal(A)\cap\mathcal(B))/ P(\mathcal(B))$, where the event
$\mathcal(B)$ consists in having fixed the time--average of the
energy, $\bar \veps=U$.

To compute the moment function we use the familiar representation for the
Dirac's function, i.e. $\int_{-L}^L \diff k
\exp(ikx)\to \delta(x)$ as $L\to\infty$. One has then
\begin{equation}
 \begin{split}    
   & Z(\lambda,A,U) =  \lim_{L\to+\infty} \int_{-L}^L \diff
   k_1\int_{-L}^L \diff k_2  
   \int_{\reali} \diff A\, \exp(-\lambda A) \\     
   & \qquad  \idotsint_{A=\frac 1N\sum A_j n_j}\diff F(n_1)\cdots
                      \diff F(n_K)\, 
             \exp\Big(i(U-\frac 1N\sum \veps_j
                       n_j)k_1+i(N - \sum n_j)k_2\Big)\\
   & \quad =  \lim_{L\to+\infty} \int_{-L}^L \diff k_1\int_{-L}^L \diff k_2 \, 
        \exp(ik_1U+ik_2N) \\
   & \idotsint \diff F(n_1)\cdots\diff F(n_K) 
     \prod_{j} \exp\Big( - n_j (\frac {\lambda A_j}N+ i\frac
        {\veps_jk_1}N +ik_2)\Big) \ .
  \end{split}        \nonumber 
\end{equation}
Notice that in the last integral there appears the Laplace 
transform of the distribution function $F(n)$.
Now we make the following
\begin{Ipotesi}
Defining $\chi(z)$ by
$$
\int_0^{+\infty} \exp(-zn) \diff F(n) =\exp(\chi(z)) \ ,
$$ 
we suppose that $\chi(z)$ is analytic in the half plane $\Re z>0$
(this is true if the random variables $n_j$ are 
supposed to be infinitely divisible).
\end{Ipotesi}
Then one has
\begin{multline}\label{eq3}
Z(\lambda,A,U) =                \\
  \lim_{L\to+\infty} \int_{-L}^L \diff k_1 \int_{-L}^L \diff k_2\,
    \exp\left(ik_1U+ik_2N+\sum_{j} \chi\Big(\frac {\lambda A_j}N+ i\frac
     {\veps_jk_1}N +ik_2\Big)\right) \ .
\end{multline}

An asymptotic
expression for the integral (\ref{eq3}) can be given through the
steepest  descent method. One has to find the values of $k_1$
and $k_2$ which satisfy the system
\begin{equation}\label{eq5}
\left\{  
      \begin{array}{lcr}
          U & = & - \frac 1N \sum_j \veps_j \chi'\Big(\frac {\lambda A_j}N+ i\frac
                       {\veps_jk_1}N +ik_2 \Big)  \\
          N & = & -   \sum_j \chi'\Big(\frac {\lambda A_j}N+ i\frac
                       {\veps_jk_1}N +ik_2 \Big)  \ .  
      \end{array}
\right.
\end{equation}
One can show that such values exist provided one makes the further 
assumption (to which we plan to come back in the future)
$$
U \le \; <\bar \veps> \ .
$$ 
It then turns out that such values are imaginary, 
so we denote them by $ik_1=\theta$ and $ik_2=\alpha$. 
In addition, one also has  to compute the determinant
$\det H(\theta,\alpha)$ of the second derivatives, which gives
\begin{equation}\label{eq4b}              
 \det H(\theta,\alpha)= \sum_{j,k} \frac {\veps_k^2-\veps_j\veps_k}{N^2} 
             \chi''\left( \frac {\lambda A_k}{N} + \frac {\theta\veps_k}{N} 
              +\alpha \right) \chi''\left( \frac {\lambda A_j}{N} 
              + \frac {\theta\veps_j}{N}+\alpha \right) \ .
\end{equation}
One has thus
\begin{equation}\label{eq8}
  Z(\lambda,A,U)\simeq \frac {\exp\left(\theta U+\alpha N 
                 +\sum_j \chi\left( \frac {\lambda A_j}{N} 
                  + \frac {\theta\veps_j}{N}+\alpha \right) \right) }
               {\sqrt{\det H(\theta,\alpha)}} \ .
\end{equation}

Obviously, to obtain  bounds on the errors, one should give an
estimate of the quantity  $\det H(\theta,\alpha)$, but this depends
on the functions $\veps$ and $\chi(z)$. 
From now on, we suppose that the denominator
in (\ref{eq8}) can be neglected in the computations of the expectations.
This will be checked to be true, in all the examples to be considered below. 
Thus, for the expectation one has the formula
\begin{equation*}
\begin{split} 
 <\bar A>_U &= -\partial_{\lambda} \log Z(\lambda,A,U)|_{\lambda=0} \\
   & = -\left( \partial_{\lambda}\theta U+\partial_{\lambda}\alpha N 
                 +\sum_j \chi'\left( \frac {\theta\veps_j}{N}+\alpha
               \right) \left( \frac {A_j}{N} + \frac 
                {\partial_{\lambda}\theta\veps_j}{N}
               +\partial_{\lambda}\alpha\right) \right)_{\lambda=0}
 \end{split} 
\end{equation*}
which finally, using (\ref{eq5}), becomes
\begin{equation}\label{eq6}
<\bar A>_U = - \frac 1N \sum_j A_j\chi'\left( \frac{\theta\veps_j}{N} + 
               \alpha \right) \ .
\end{equation}
With this result, the form of the expressions (\ref{eq5}) and 
(\ref{eq6})  shows that
the quantity $-\chi'(\veps_j\theta/N+\alpha)$ plays the role of the mean
number of times the cell $\Zi$ is visited. Indeed, in terms of the
quantities 
\begin{equation}\label{eq6b}
\bar \nu_j\=-\chi'(\veps_j\theta/N+\alpha) \ ,
\end{equation} 
such relations can be
written as $N=\sum \bar \nu_j$, $U=\sum \veps_j \bar \nu_j/N$ and  
$<\bar A>_U=\sum A_j\bar \nu_j/N$. The quantity $\bar\nu_j$ may 
be called the mean occupation number of cell $\Zi$.

\section{The Thermodynamics}

The expression (\ref{eq6}) solves the problem of computing the conditional 
expectation, but one can ask for  the meanings
of the two quantities $\theta$ and $\alpha$. Let us consider the
problem of $\theta$. 
To this end it is convenient 
to introduce as an indipendent variable, instead of $z_j$, the quantity
$\nu_j=-\chi'(z_j)$,  and this naturally leads to introducing 
in place of $\chi$ its Legendre transform $h$ defined as  usual by
$$
h(\nu_j)=\big( \nu_jz_j+\chi(z_j)\big)\big|_{\nu_j=-\chi'(z_j)} \ .
$$
Notice that, while $\bar \nu_j$ has the meaning of a mean occupation
number (conditioned on $U$), the quantity $\nu_j$ just plays the role
of a parameter, in the same sense as $z_j$ does. In particular, the
quantities  $\nu_j$ do not need satisfy any condition related to
normalization, or the fixing of an energy value. 
One has then
\begin{Proposizione}
The values $\bar \nu_j =-\chi'(\veps_j\theta/N+\alpha)$,
$i=1,\ldots,K$, correspond to a maximum of the function
\begin{equation}\label{eq11}
  S(\nu_1,\ldots,\nu_K)\=\sum_{j=1}^K h(\nu_j)\,\ ,
\end{equation}
constrained to the surfaces $\sum \nu_j =N$ and $\sum \veps_j\nu_j/N=U$.  
\end{Proposizione}

\vskip .5 truecm
\noindent
\textbf{Proof.} One simply  considers the function
$G\=S-(\theta/N)\sum\veps_j\nu_j - \alpha\sum \nu_j$, and from
$\partial_{\nu_j}G=0$ one gets
\begin{equation}\label{eq7}
h'(\bar \nu_j)= \frac {\theta\veps_j}{N} + \alpha \ .
\end{equation}
But now, from the Legendre duality, one has
$$
h'(\bar \nu_j)= \frac {\theta\veps_j}{N} + \alpha 
\qquad \Longleftrightarrow \qquad
\bar \nu_j    = -\chi'\left(\frac {\theta\veps_j}{N} + \alpha\right) \ ,
$$
from which the thesis follows. One easily checks that one actually
deals with a maximum. Q.E.D.

\vskip .5 truecm
\noindent
As a corollary there follows that the maximum of $S$ (divided by $N$) is
indeed the thermodynamic entropy,  and $\theta/N$ the inverse
temperature, or  at least an integrating factor of the exchanged heat. 
In other terms, the thermodynamic entropy
$S^{\mathrm{th}}(\theta,\alpha)$
turns out to be given by
$$
S^{\mathrm{th}}(\theta,\alpha)
\= S(\bar\nu_1,\ldots,\bar\nu_K)/N \ ,
$$ 
up to an additive constant.

In fact, suppose now that the values $\veps_j$ depend on some external
parameter, say $\kappa$. Then $\partial_{\kappa}\veps_j$ is the
reaction force needed to keep the parameter fixed when the system
is in cell $\Zi$ of the phase space, so that the quantity
$\partial_{\kappa}\veps_j\diff\kappa$ is the (instantaneous) work
performed on the system when  the external parameter is changed by a
quantity $\diff\kappa$. For the macroscopic work $\delta
\mathcal{W}$, namely the expectation of the time--averaged
instantantaneous work, using relation (\ref{eq6}) 
one then obtains the expression
$$
\delta \mathcal{W} = \frac 1N \sum_{j} \bar\nu_j
\partial_{\kappa}\veps_j\diff\kappa \ .
$$
By definition, the exchanged heat is then
\begin{equation*}
  \begin{split}
    \delta Q &\= \diff U -\delta \mathcal{W}            \\
             & = \diff\left( \frac 1N \sum_{j} \bar \nu_j\veps_j \right) - 
                 \frac 1N \sum_{j} \bar\nu_j 
                 \partial_{\kappa}\veps_j\diff\kappa     \\
        & = \frac 1N\sum_{i} \veps_j\diff \bar\nu_j \ .
  \end{split}
\end{equation*}
On the other hand, from (\ref{eq7}) one has $\veps_j/N=(h'(\bar
\nu_j)-\alpha)/\theta$, so that one finds
\begin{equation*}
  \begin{split}
    \delta Q & =  \frac 1{\theta}\sum_{i}h'(\bar\nu_j) \diff \bar\nu_j - 
          \frac {\alpha}{\theta} \sum_j \diff \bar \nu_j \\
   & =  \frac N{\theta} \diff \left( \frac 1N \sum_{j}h(\bar\nu_j) \right)
              = \frac N{\theta} \diff S^{\mathrm{th}} \ ,
  \end{split}
\end{equation*}
because in our hypotheses $N$ is kept constant, so that
$\sum\diff\bar\nu_j=0$.  Concerning the physical meaning of $\alpha$,
we have no clear idea at the moment.

\section{The Gibbs distribution}
The most natural choice for the distribution function that one might
consider is that of  Poisson, namely
$$
F(n_j)= \sum_{k\le n_j}\frac {e^{-p}}{k!} p^{k} \ ,
$$
where $p$ is a positive parameter.
This corresponds to assuming that the successive visits of a given cell 
are independent events. 
 
The Laplace transform then  has the form
$$
\int_0^{+\infty} e^{-nz}\diff F = e^{-p}\,  \sum_{n=0}^{+\infty} e^{-nz}
     \frac {p^n}{n!}=\exp(pe^{-z}-p) \ ,
$$
which in particular exhibits the well known fact that the distribution 
is infinitely divisible. One has  thus
$$
\chi(z)=pe^{-z}-p \ ,
$$
and 
$$
\bar \nu_j = -\chi'(\frac {\theta \veps_j}{N}+\alpha)
         = pe^{-\alpha}e^{-\theta \veps_j/N} \ .
$$
The condition $\sum \bar \nu_j=N$ then gives $pe^{-\alpha}=N/Z(\beta)$,
where $Z(\beta)\=\sum  e^{-\beta \veps_j}$ 
is the usual canonical partition function, and
$\beta=\theta/N$. Thus, for the mean occupation number 
one has the usual Gibbs formula
$$
\frac {\bar \nu_j}{N} =\frac {e^{-\beta \veps_j}}{Z(\beta)} \quad .
$$

With this result,  the value of the denominator in (\ref{eq8}) 
can be computed, and  one gets
\begin{multline}
\sum_{j,l}  \frac {\veps_l^2-\veps_j\veps_l}{N^2} 
        \chi''\left( \frac {\lambda A_l}{N} + \frac {\theta\veps_l}{N} 
        +\alpha \right) \chi''\left( \frac {\lambda A_j}{N} 
        + \frac {\theta\veps_j}{N}+\alpha \right)= \\
 = \sum_j \frac {\veps_j^2e^{-(\lambda A_j/N)-\beta\veps_j}}
        {Z(\beta)} - \left(\sum_j \frac {\veps_je^{-(\lambda A_j/N)
        -\beta\veps_j}}{Z(\beta)} \right)^2 \ ,\nonumber
\end{multline}
a quantity which tends, for $\lambda\to0$, to the familiar expression
for the canonical specific heat $C_V$. So, 
if the specific heat $C_V$ is an extensive
quantity of the same order of magnitude as $U$, as usual for the 
systems dealt with in statistical mechanics, it follows that the steepest
descent method gives the correct answer. In addition,
in computing
the expectation of $\bar A$ the denominator gives a contribution
equal to
$$
\partial_{\lambda} \log C_V=\frac {\partial_\lambda C_V}{C_V} \ ,
$$
which is of order $O(1)$ for quantities $A$ of macroscopic type, 
and so can be safely ignored. 

Coming finally to the computation of the entropy, one first has to compute
the quantity $h(\nu)=\nu z+pe^{-z}-p$ using  $\nu=pe^{-z}$, i.e. 
$z=-\log \nu+\log p$, so that one has $h(\nu)=-\chi'(z)=-\nu\log \nu 
+ \nu + \nu\log p-p$. 
Thus for the entropy one finds the expression
$$
S=-\sum_j \nu_j \log \nu_j + N(1+\log p) - Kp \ ,
$$ 
i.e., up to an additive constant, 
the classical expression of Boltzmann (or rather the analogous one  of Gibbs involving the global phase--space).

\section{ The infinitely divisible distributions and the Tsallis  entropies}
The infinitely divisible distributions are
characterized according to the following theorem (see the classical
handbook \cite{feller} of Feller)
\begin{Teorema} 
The function $\exp(\chi(z))$ is the Laplace
transform of an infinitely divisible distribution if and only if
$\chi(0)=0$ and $(-1)^n\chi^{(n)}(z)>0$, $\forall z\in\reali^+$ and
$\forall n\in \interi$.
\end{Teorema}
Perhaps, the simplest choice consists in considering the
inverse  powers of $z$. 
Taking into account normalization conditions, one is then naturally led
to consider the one--parameter family,
parameterized by $\gamma>0$, which, for any positive constant $p$, 
is given by
\begin{equation}\label{eq9}
\chi(z)=p\left(1+\frac {z}{\gamma}\right)^{-\gamma} - p 
\ .
\end{equation}
Notice that it reduces to the Poisson distribution 
in the limit $\gamma\to+\infty$. 

The mean occupation number $\bar \nu_j$ is then given,
using  (\ref{eq6b}), by
$$
\bar \nu_j = p\left(1+\frac {\theta\veps_j}{N\gamma} +
           \frac {\alpha}{\gamma}\right)^{-\gamma-1}
\ .
$$
Now, it turns out that 
this coincides, with a suitable relabeling of the parameters, 
with the Tsallis distribution of index $q$ 
(see \cite{tzallis}),
$$
\bar \nu_j = C(1+\beta_q(q-1)\veps_j)^{\frac {q}{1-q}}
$$
if one sets 
$$
\gamma = \frac 1{q-1} \ .
$$
Notice that, in terms of the
parameter $q$, the one--parameter family (\ref{eq9}) 
can also be written in a form reminiscent of the Poisson distribution, namely
$$
\chi(z)=pe_q^{-z}-p \ ,
$$
where $e_q^z$ is the $q$ deformation of the exponential introduced by
Tsallis, namely
$$
e_q^x=(1+(1-q)x)^{\frac 1{1-q}} \ .
$$ 

The fact that the mean occupation number, as a function of energy, is
a $q$--distribution 
is due to the fact that the entropy (\ref{eq11})
corresponding to distribution (\ref{eq9}) 
essentially coincides with that of Tsallis. 
In fact, from (\ref{eq9}) one has
$$
\nu_j=p\left(1+\frac {z_j}{\gamma}\right)^{-\gamma-1} \ ,
$$
so that, computing $z_j$ as a function of $\nu_j$, one gets
$$
h(\nu_j) = (\gamma +1)\nu_j^{\frac {\gamma}{\gamma+1}} p^{\frac
           1{\gamma+1}} -\gamma \nu_j -p\ .
$$ 

Apparently, the entropy $S=\sum h(\nu_j)$ 
thus obtained does not have the form of the Tsallis
entropy, even if $\gamma$ is expressed in terms of $q$ as above.
However the coincidence (apart from an inessential multiplicative
constant) is obtained if the
present entropy is written in terms of the quantities 
$p_j\=p^{1/q}q^{1/q-1}\nu_j^{1/q}$ and if the mean energy is 
computed in terms of ``escort'' probabilities (see \cite{beck}), 
namely $p_j^q$, i.e. essentially our $\bar \nu_j$.
Indeed, in terms of the  variables $p_j$ and  of the parameter $q$ the
present entropy $S$ reads 
$$
S(p_1,\ldots,p_K) = C(q) \frac { \sum_j p_j - \sum_j p_j^q}{q-1} \ ,
$$
where we have introduced the constant 
$C(q)\=pq^{(q-1)/q}$.
This expression differs from that of Tsallis  
only by a multiplicative constant, so that obviuosly both entropies 
produce by maximization the same distribution.

In a such a way the expressions of the Tsallis entropies are
recovered. The main difference seems to be that the present
procedure, in virtue of its statistical--mechanics foundation, does not
require any jumping between probabilities and escort probabilities.

\section{Conclusions} 
We believe we have given a consistent procedure to define a
thermodynamic entropy for off--equilibrium situations. 
A characteristic feature of the present approach is that it follows
the same scheme of Gibbs and Khinchin, which consists  in finding an integrating
factor for the exchanged heat, with no mention of information theory
at all. We are well aware however of the fact that another problem
remains open, because in general there exist
infinitely many integrating factors, so that a
further requirement is needed in order to uniquely determine the thermodynamic
entropy. This fact was particularly emphasized for example by Ehrenfest and 
Caratheodory (see \cite{ehrenfest}). Notice that such a requirement
involves the question of the extensivity property of entropy. 
These interesting problems are left for possible future studies.
Another subject for further study is the dependence of the entropy on
the number $N$ of iterations, i.e. on time.

The second peculiarity of the present approach concerns the
essential ingredient through which probability was introduced, namely the
probability distribution $F(n_j)$ for the number of visits of cell $\Zi$,
against the  common use of interpreting  
the numbers $n_j/N$ themselves somehow as probabilities.
It has been shown how  thermodynamics turns out to be defined in terms of the
function $\chi(z)$ uniquely associated to the
distribution $F(n_j)$. 
Notice that the treatment developed here can be easily extended to the
nonhomogeneous case, in which the probability distribution $F$ depends
on cell $\Zi$. In such a case the function $\chi(z)$ too depends
on $j$ as does the function $h$. One has then for example
$\bar\nu_j=-\chi_j'(\veps_j\theta/N+\alpha)$ and
$S(\nu_1,\ldots,\nu_K)=\sum h_j(\nu_j)$, and everything  remains
essentially unchanged.

\end{document}